\documentclass{elsart}

 \usepackage{graphicx}
 \usepackage{epsfig}
\usepackage{amssymb}
\usepackage[latin1]{inputenc}
\usepackage{graphicx}  
\usepackage{bm}        
\usepackage{latexsym}
\usepackage{mathrsfs}
\usepackage{amssymb}
\usepackage{amsmath}
\usepackage{amscd}
\usepackage{color}
\usepackage{verbatim}
\journal{Physica D}

\begin{document}
\centerline{\bf \large Accepted Physica D}

\begin{frontmatter}



\title{Chaotic motion at the emergence of the time averaged energy decay}
\author{C.~Manchein},
\author{J.~Rosa\thanksref{cor2}},
\author{M.~W.~Beims\corauthref{cor}}
\thanks[cor2]{Permanent address: Escola Técnica Federal, 
  Unidade de Paranaguá, 83215-750, Paranaguá, PR, Brazil}
\corauth[cor]{Corresponding author}
\ead{mbeims@fisica.ufpr.br}

\address{Departamento de Física, Universidade Federal do Paraná, 81531-990,
  Curitiba, PR, Brazil.}

\begin{abstract}
A system plus environment conservative model is used to characterize the
nonlinear dynamics when the time averaged energy for the system particle 
starts to decay.  The system particle dynamics is regular for low values 
of the $N$ environment oscillators and becomes chaotic in the interval  
$13\le N\le15$, where the system time averaged energy starts to decay.
To characterize the nonlinear motion we estimate the Lyapunov exponent 
(LE), determine the power spectrum and the Kaplan-Yorke dimension.
For much larger values of $N$ the energy of the system particle is 
completely transferred to the environment and the corresponding LEs 
decrease.
Numerical evidences show the connection between the variations of the 
{\it amplitude} of the particles energy time oscillation with the time
averaged energy decay and trapped trajectories.
\end{abstract}

\end{frontmatter}

\section{Introduction}
\label{intro}
Small dissipation is inevitable in real systems. 
In the Langevin description \cite{lindenberg,risken} the damping force,
together with the fluctuating force per unit mass (Langevin force), are used 
to model dissipation and friction. The damping force (or friction) 
is given by the Stoke's law $F_s=-\gamma v$ and $\gamma$ is the damping 
constant. These forces represent the effect of the collisions of the 
system particle with the bath particles and are
related via the dissipation-fluctuation theorem \cite{risken}. In the Langevin
description we deal with a one-dimensional differential equation with
dissipation.
For higher-dimensional systems the whole dynamics follows the properties of 
many coexisting attractors. The characterization of this dynamics is not 
simple since high-dimensional systems usually present all characteristics of 
complexity \cite{grebogi96}, which may arise in the interplay between small 
dissipation and a conservative dynamics \cite{grebogi96,rech05}. 
Even for low-dimensional systems, elliptic periodic orbits become small sinks 
when small dissipation is added and an infinite number of periodic attractors 
may coexist \cite{feudel96}. Therefore, a nonlinear analysis on the borderline 
between small dissipation and a conservative dynamics is of interest
for the description of complex systems in the real world. 

Another possibility to describe dissipation is to consider an {\it open system} 
interacting with its {\it environment} by collision processes.  The whole 
problem (System $+$ Environment $+$ Interaction) is conservative but, due to
energy exchange between system and environment, the {\it system} can be 
interpreted as an open system with dissipation. Such a microscopic theoretical 
model has been proposed to describe dissipation processes in quantum  
\cite{caldeira,weiss,classical} and in classical systems 
\cite{lindenberg,classical}. 
In the limit of an infinite number of environment constituents, it is 
possible, in some specific cases, to derive a Langevin equation from the 
microscopic model \cite{lindenberg}. Different from these works,
in this paper the environment is composed of a {\it finite} number $N$ of 
uncoupled harmonic oscillators.  Changing $N$ we are able to study the
interesting  transition from  low- to high-dimensional dynamical
systems which simultaneously experiment dissipation.
The microscopic dynamics in this transition is not simple and its relation
to macroscopic statistical mechanics is even more complicated. Such 
transitions started to be studied in the context of the Fermi-Pasta-Ulam 
(FPU) model \cite{fermi} in order to answer questions 
related to irreversible statistical mechanics and equipartition of energy. 
The FPU model is a  one-dimensional chain of $(N-1)$ coupled moving mass 
points (See Ford \cite{ford} for a nice review). The main problem was to 
explain the non-equipartition of energy observed for high values of $N$. 
With the help of chaos theory it became clear later that the equipartition
of energy can be achieved at the strong stochasticity threshold 
\cite{radons}.
In the model considered here the finite $N$ harmonic oscillators are 
{\it indirectly} coupled via the system particle. It is therefore a 
different physical situation from the FPU model and from the chaotic 
two-dimensional systems used to model thermal baths \cite{aguiar}. Beside 
that, it has been 
shown recently \cite{Jane3} that for finite values of $N$ the environment 
induces a non-Markovian motion on the system particle, and consequently, 
non-exponential energy decays. Therefore, our reservoir induces a damping
mechanism which, in the Langevin description, corresponds to a complicated
memory kernel \cite{lindenberg} which only recovers the usual Stokes friction
$\gamma$ in the limit $N\rightarrow\infty$ and for high environment 
frequencies (for details see \cite{Jane3}). In fact, our friction
is determined during the time evolution through the dynamical process. 
Further, our damping mechanism differs from the velocity dependent friction 
in the Gaussian thermostat, obtained the requiring energy conservation at
any time, and from the more general damping mechanism proposed in the 
Nos\'e-Hoover thermostat \cite{klages}. The finite $N$ model considered here 
has been used to study environment effects on the transport of particles in 
ratchets \cite{Jane1} and the role of Levy walks on the mobility in ratchet
potentials \cite{Jane2}.

In this paper we focus on the nonlinear dynamics of the time series obtained
for the system particle under the $N$ harmonic oscillators. The tools used 
in this analysis are the Lyapunov spectrum \cite{sano}, phase space dynamics,
power spectrum and the Kaplan-Yorke dimension. Our system is deterministic 
and we use the TISEAN \cite{tisean} package to analyse the time series 
\cite{tis}. The paper is organized as follows. Section \ref{model} presents 
the model and Section \ref{dissi} discusses the values of $N$ where the time 
averaged system energy starts to decay. In  Section \ref{results} results for 
the nonlinear 
time series analysis for one initial condition are presented. Lyapunov 
exponents, power spectrum and the Kaplan-York dimension are discussed. Finally 
in Section \ref{conclusion} the conclusions of the paper are presented.

\section{The Model}
\label{model}

\begin{figure}[htb]
 \begin{center}
\includegraphics*[width=8.5cm,angle=0]{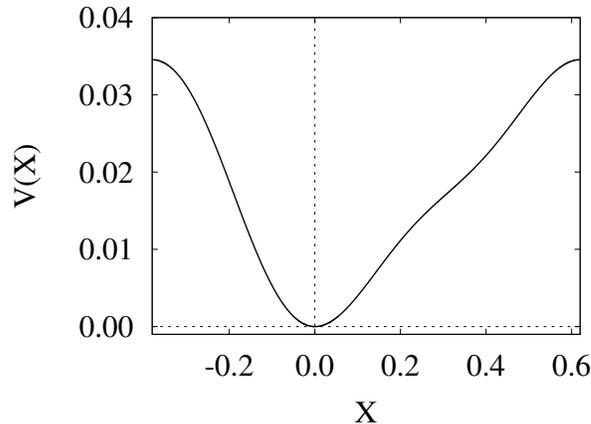}
 \end{center}
 \caption{Dimensionless anharmonic potential $V(X)$ for $-0.38\le X\le 0.62$.}
  \label{anhar}
  \end{figure}

Let us consider the problem composed by a particle under the influence of an
asymmetric potential (the system) interacting with $N$ independent harmonic 
oscillators (the environment). The dimensionless equations of motion are (see 
\cite{Jane2} for more details)

\begin{equation}
\ddot{X} + \frac{dV(X)}{dX}-\sum_j^N{\gamma_jx_j}
+X \sum_j^N{\frac{\gamma_j^2}{\mu_jw^2_j}}
=0
\textit{,}
\label{eq:mov4}
\end{equation}

\begin{equation}
  \ddot{x}_j+{w^2_j}x_j-\frac{\gamma_j}{\mu_j}X=0
~\textit{,}
   \label{eq:mov5}
\end{equation}
where $X$ and ${\vec x}$ are, respectively, system and oscillators
$(x_j, j=1,2,...,N)$ coordinates. The coupling parameter $\gamma_j$ is
a measure of the strength of the bilinear coupling between the system
and the $j$-oscillator. The dimensionless anharmonic potential (see 
Fig.~\ref{anhar}) is defined in the interval $X=(-0.38,0.62)$ by


\begin{displaymath}
  V(X)=C-\frac{1}{4\pi^2\delta}\left[\sin{2\pi(X-X_0)}+\frac{1}{4}
\sin{4\pi(X-X_0)}\right] ~\textit{.}
\end{displaymath}\\
%
The constant $C$ is such that $V(0)=0$. The time and the oscillators 
frequencies $w_j$ are written in units of  $w_0=1.0$, which is the frequency of 
the linear motion around the minimum of the potential and is determined from 
$w^2_0={4\pi^2V_0\delta}/{M}$, where $\delta=\sin{2\pi|X'_0|}+\sin{4\pi|X'_0|}$.
The dimensionless equations of motion (\ref{eq:mov4}) and (\ref{eq:mov5}) are 
the equations analyzed in this work. We used fourth-order Runge-Kutta integrator
\cite{numeric} with fixed step $\Delta t=10^{-3}$.

Some words about the model are in order. Although we solve all $2N+2$ 
equations of motion, the main dynamics is observed by the {\it system}
particle. Usually in such system-plus-environment models with a bilinear
coupling, the coordinates of the environment can be eliminated in the 
limit $N\rightarrow\infty$ and the resulting equation of motion for the 
system particle is a kind of Langevin equation \cite{lindenberg,weiss}. 
To notice in such approach is the assumption of weak coupling between 
the system and environment, and the {\it infinite} number of oscillators 
of the environment which characterize a thermal bath which is not affected
by the system dynamics. Since in this work we investigate numerically
the full set of equations of motion (\ref{eq:mov4}) and (\ref{eq:mov5}), 
the energy exchange and energy decay can be analyzed explicitly by 
increasing, one by one, the values of $N$.

For all cases considered in this paper, dimensionless mass $\mu_j=0.1$ and
$\gamma_j=0.01$ are used. At time $t=0$, the system and the environment are not 
considered to be in equilibrium. The interaction energy is assumed to be 
zero, the energy of the system is very close to the total energy 
$E_S\sim E_T=0.02$, and the oscillators energy is close to zero 
($E_O\sim 0.0$). The energy in all figures is adimensional. The initial 
distribution of the oscillators position and velocity is uniform around zero.
For high values of $N$ the distribution of the oscillators frequencies 
approaches a quadratic (Debye-type) distribution with a cutoff at 
$w_{\mbox{cut}}=2.1$. The lower limit of this distribution is $w=1.1$.

In this work we kept the total energy $E_T$ fixed instead of the
energy density $\varepsilon=E_T/N$. The reason to do this is simple. Our
goal is to analyze the nonlinear dynamics of the system particle when 
energy is transferred from the system to the environment when 
$N=1\rightarrow4000$. If we hold $\varepsilon$ fixed, two limiting
situations are obtained. In the first one (when $N\rightarrow4000$) 
we must increase $E_T$ and the system particle is not initially inside the 
potential well of Fig.~(\ref{anhar}). In such situation the particle will 
need a long time to loose its energy (via collisions with the $N$ 
oscillators) until it ``falls down'' into the potential well. When this 
happens, the physical situation is almost identical to the case discussed 
in this paper, where the particle starts inside the potential well. In the 
other limit (when $N=1$) the energy $E_T$ must be reduced very much and 
no energy transfer can be observed. In fact, both analysis could be
implemented, but for the purpose of the present paper it is more appropriate
to keep the total energy fixed.

\section{Emergence of the time averaged energy decay}
\label{dissi}

The system particle exchanges energy with the $N$ environment oscillators.
This energy, continuously transferred to the environment, can return back to 
the system particle after a time of the order of the Poincar\'e recurrence 
time \cite{weiss}. In general, for lower values of
$N\lesssim15$, the recurrence times are not very large and the energy return 
can be observed in simulations. However, for higher values of $N$, the 
Poincar\'e recurrence times increase very much to be observable with finite 
integration times. In such cases we can say that the energy transferred to 
the environment will not return, within the integration times, to the system 
particle. This is shown in 
 \begin{figure}[htb]
 \begin{center}
 \includegraphics*[width=8cm,angle=0]{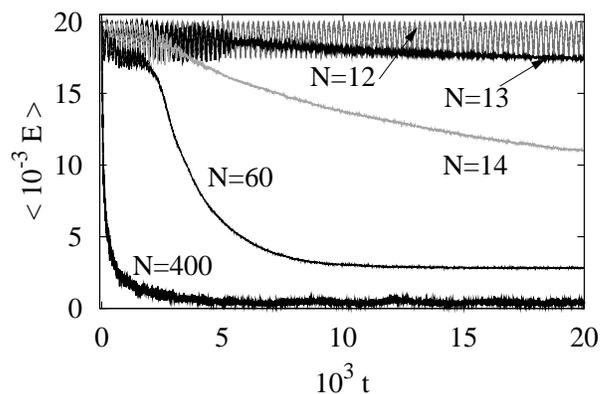}
 \end{center}
 \caption{Mean energy (over $800$ environmental initial conditions) as a 
          function of time for $N=12,13,14,60,400$.}
  \label{En-N12-15} 
  \end{figure}
Fig.~\ref{En-N12-15}, where the mean normalized system energy is plotted
as a function of time for $N=12,13,14,60$ and $400$.  For $N=12$ this 
energy oscillates strongly around the mean value $\sim 0.019$, but its 
time average is constant. For $N=13$ we see that, besides initial strong 
oscillations, the time averaged energy starts to decrease at times 
$t_{13}\sim 5\times10^3$. This means that at times $t\sim t_{13}$ a portion 
of the system particle energy starts to be transferred to the environment
without returning. For $N=14$ the time averaged energy starts to decay  
at times $t_{14}\sim3\times10^3$ and decreases continuously for the whole 
range of integration times. Increasing the values of $N$ to $60$ and $400$, 
we observe that the time at which the mean energy starts to be transferred 
to the environment gets smaller. 
It is interesting to observe that the particle energy starts to be
transferred to the environment exactly when the {\it amplitude} of the 
large time oscillations of the mean energy decreases. 

An interesting characteristic to mention from Fig.~\ref{En-N12-15} is the 
non-exponential energy decay. For example, for $N=13$ the fit for the 
energy obeys $\left< E\right>\sim0.029\  t^{-0.052}$ while for $N=400$ it 
obeys $\left< E\right>\sim0.275\ t^{-0.733}$. Such power law behaviour is 
related to the non-Markovian dynamics induced by the environment and can 
be explained via environment autocorrelation function (see \cite{Jane3} 
for details.)

 \begin{figure}[htb]
 \begin{center}
 \leavevmode
 \includegraphics*[width=8cm,angle=0]{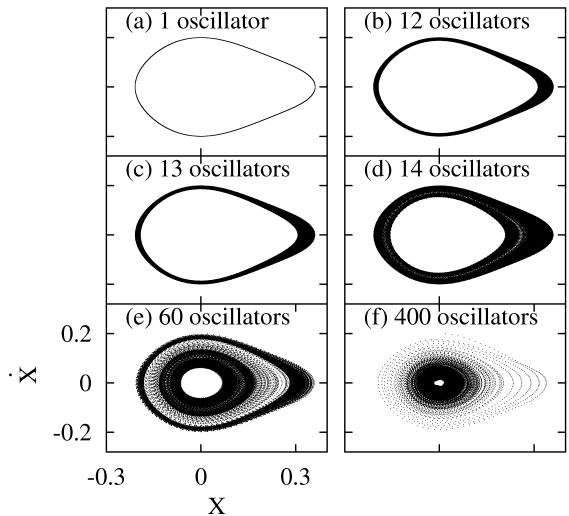}
 \end{center}
 \caption{Phase space dynamics of the system particle for 
  $N=1,12,13,14,60,400$.}
  \label{PS-N12-15} 
  \end{figure}
Figure \ref{PS-N12-15} shows the phase space dynamics of the system 
particle for $N=1,12,13,14,60,400$. For $N=1$ the dynamics is close 
to a deformed elliptic motion. As $N$ increases to $N=12,13,14$,
the system particle exchanges energy with the oscillators and moves 
within a layer inside the deformed ellipsis. This layer increases as 
$N$ increases. For higher values of $N$ ($60$ and mainly $400$) the system 
particle rapidly loses its energy to the environment, and ends up moving 
close to the minimum of the anharmonic potential. When $N=4000$ (not shown) 
the system particle energy is close to zero and the energy per oscillator
is also close to zero. 

 \begin{figure}[htb]
 \begin{center}
\includegraphics*[width=14cm,angle=0]{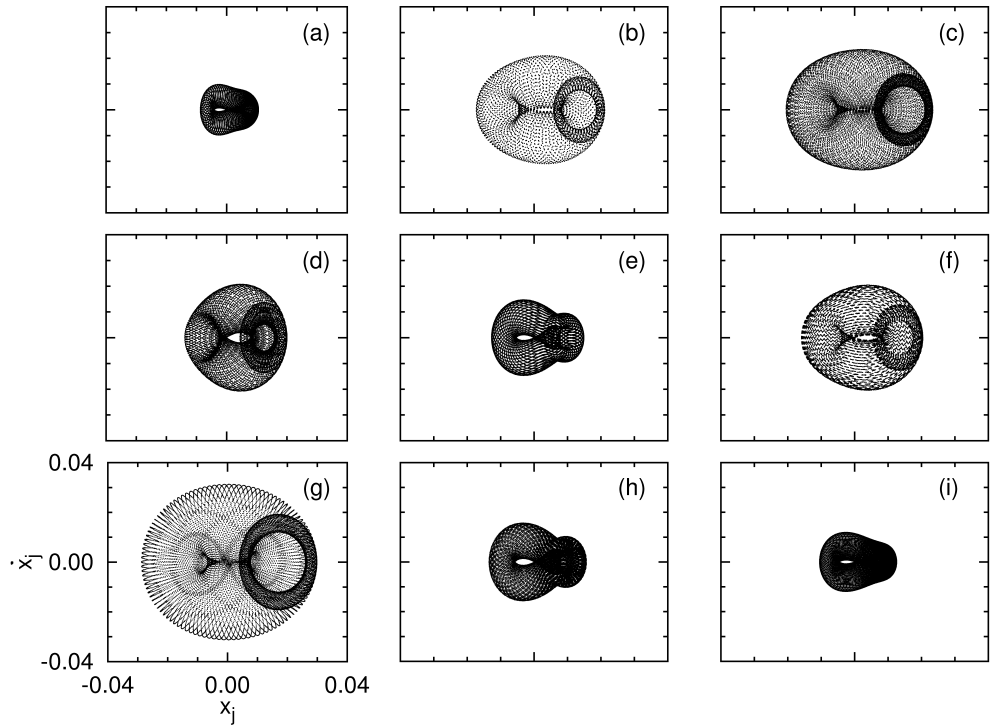}
 \end{center}
 \caption{Phase space dynamics of the environment oscillators for $N=9$.
      From left to right and from top to bottom. The oscillators 
     frequencies $w_j$ are
     (a) $2.089$, (b)$1.594$, (c) $1.529$, (d) $1.669$, (e) $1.875$, 
     (f) $1.625$, (g) $1.414$, (h) $1.853$, (i) $1.969$.}
  \label{PS9} 
  \end{figure}
 \begin{figure}[htb]
 \begin{center}
\includegraphics*[width=14cm,angle=0]{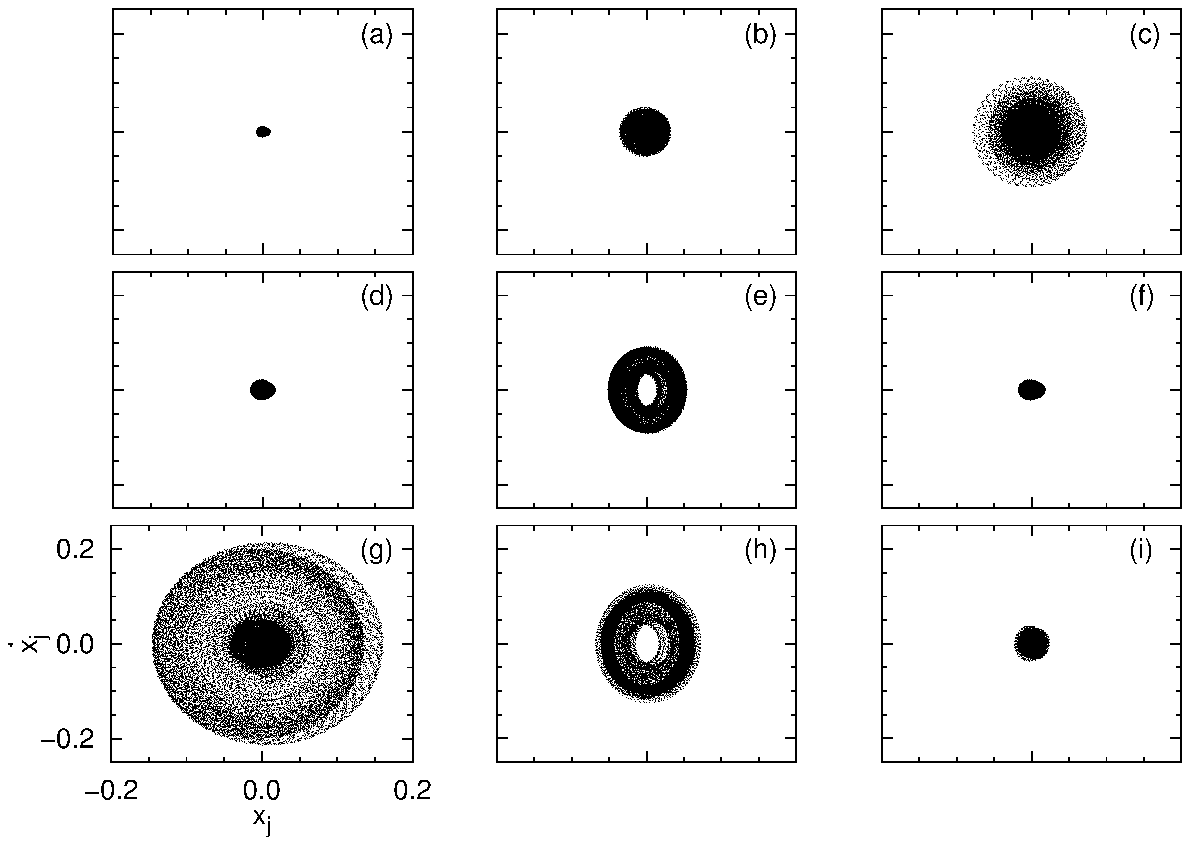}
 \end{center}
 \caption{Phase space dynamics for the first nine environment oscillators 
       for $N=15$. From left to right and from top to bottom. The frequencies
       are given in the caption of Fig.~\ref{PS9}.}
  \label{PS15} 
  \end{figure}

Figures \ref{PS9} and \ref{PS15} show respectively the phase space 
dynamics for the environment oscillators for $N=9$ and $N=15$.
The oscillators are shown from left to right and from top to bottom. 
For $N=15$ we only display the first nine oscillators. For $N=9$
the shape of the curves in phase space are essentially governed by 
the frequency $w_j$ of each oscillator. The magnitudes of the 
oscillations in phase space (the size of the curves) are inversely
proportional to the frequencies $w_j$. Environment oscillators with 
lower frequencies are more likely to receive the system particle energy 
and, consequently, the size of the oscillations increase. See for example 
the size of  the lowest frequencies oscillations from 
Figs.~\ref{PS9}(c) and (g). For oscillators with higher frequencies 
the size of the oscillations is smaller, as can be observed in 
Figs.~\ref{PS9}(a) and (i). We essentially observe (together with 
conclusions obtained from 
Fig.~\ref{En-N12-15}) that for $N=9$ the system and oscillators
just exchange energy continuously and the geometrical curves in
phase space look ``well behaved''. The shape of these curves does
not change in time and consequently no equipartition of energy 
is observed for the integration times. For $N=15$ (see Fig.~\ref{PS15}), 
when the time averaged energy decay has already occurred, all ``well 
behaved'' curves from the oscillators start to be destroyed and 
change in time. For small times (not shown) they look very similar 
to the curves from $N=9$, but for later times ($t>1.5\times 10^3$) 
they converge to 
the curves displayed in Fig.~\ref{PS15}. Moreover, although energy 
is slowly shared between oscillators we observe that the magnitudes 
of oscillations in phase space are still different for distinct 
oscillators. For example, observe that the oscillation 
magnitude of case (g), with the lowest frequency, is much larger 
than the other ones (note that the scale from Fig.~\ref{PS15}
differs from the scale of Fig.~\ref{PS9}). This shows us again that 
we do not have equipartition of energy. We also determined
the oscillator phase space dynamics for $N=4000$ (not shown) and
non-equipartition of energy was observed.

Concluding this section we observe that the time averaged energy decay 
for the system particle starts to occur in the interval $N=(12,14)$. 
For lower values of $N$ the system 
continuously exchange energy with the environment but the time average is
constant. For much higher values of $N$ the system particle energy is  
transferred to the environment for times very close to zero. The 
above mentioned interval $N=(12,14)$ is valid for the specific
oscillators frequencies generator used in the simulations. These values of 
$N$ may change a little if another generator is used. We checked this result 
for other frequencies generators and observed that the time averaged energy
decay for the system particle always occurs for $N$ in the interval $(10,20)$ 
\cite{Jane3}. This is (not coincidently) approximately the interval of $N$ 
where the Poincar\'e recurrences times diverge \cite{weiss}. So, this 
interval should be typical  
when  many degrees of freedom are coupled and it should not depend strongly
on the system potential shape (or system parameters). The potential 
shape and system parameters will affect the {\it form} of the dynamics in 
phase space (Figs.~\ref{PS-N12-15}, \ref{PS9}, \ref{PS15}) and the {\it rate}
of energy decay in Fig.~\ref{En-N12-15}, but not the interval {\it where} 
the time averaged energy starts to decay.

\section{Nonlinear analysis}
\label{results}

By integrating Eqs.~(\ref{eq:mov4}) and (\ref{eq:mov5}) for many values of $N$,
we generated the time series (TS) for the variable $X(t)$ for the system 
particle only. For all simulations we used the same initial condition 
$X(0)=0.0$ and $V(0)=0.2$. The system particle has not enough energy to 
``jump'' over the potential barrier. In order to perform the nonlinear 
analysis it is necessary to determine the dimension $m$ of the reconstructed 
attractor \cite{takens}. The adequate values of $m$ were determined using the
false-nearest neighbors method \cite{kennel}. For values of $N=1\rightarrow4$, 
the appropriate embedding dimension is $m=2$, for $N=5\rightarrow12$ it is 
$m=3$, for $N=13$ and $N=14$ it is $m=4$ and for $N\ge15$ it is $m=5$. These 
are the values of $m$ used in all simulations. 

Equations (\ref{eq:mov4}) and (\ref{eq:mov5}) were integrated until 
$t=2\times10^4$. The number of points used in the time series  were  
$2\times10^5$. We start by showing results for low values of $N$, where 
the system particle exchanges energy with the $N$ environment oscillators.
Then we increase $N$ to values where the time averaged energy 
decay emerges and finally we discuss the case of high values of $N$.

\subsection{$N=0$. } This is the case of no environment. The particle
oscillates inside the anharmonic potential shown in Fig.~\ref{anhar}.
Since the phase-space is two-dimensional the motion is non-chaotic.

\subsection{$N=1\rightarrow 12$} 
\begin{figure}[htb]
 \begin{center}
 \includegraphics*[width=9cm,angle=0]{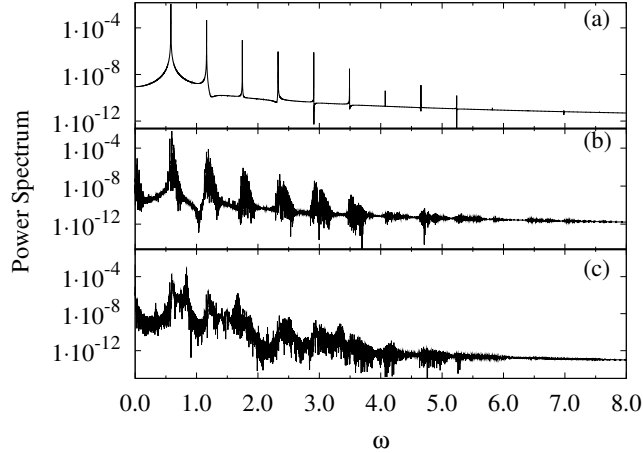}
 \end{center}
 \caption{Power spectrum for (a) $N=1$, (b) $N=13$ and (c) $N=20$.}
  \label{PS-N10} 
  \end{figure}

For $N>0$ the TS of the system particle dynamics starts to change due to 
effects coming from the environment. In the interval  $1\le N\le 12$,
all LEs  are time independent and are estimated to be $\lesssim10^{-3}$. 
Such LEs can be considered zero. Figure \ref{PS-N10}(a) shows 
the power spectrum for $N=1$. We observe that the peaks 
are well located at specific values of the frequencies. The main
frequency corresponds to the unscaled particle angular frequency 
$\omega\sim0.56$Hz, which is the fundamental frequency of the motion 
in the anharmonic potential. All other peaks are high-harmonics 
of the fundamental frequency. The power spectra for $N=2\rightarrow12$ 
(not shown) shows some small frequencies around the peaks from 
Fig.~\ref{PS-N10}(a). From this we conclude that, together with the zero 
LEs, the system particle dynamics is regular for values of $N\le12$. 

\subsection{$N=13\rightarrow20$}
\begin{figure}[htb]
 \begin{center}
\includegraphics*[width=8.5cm,angle=0]{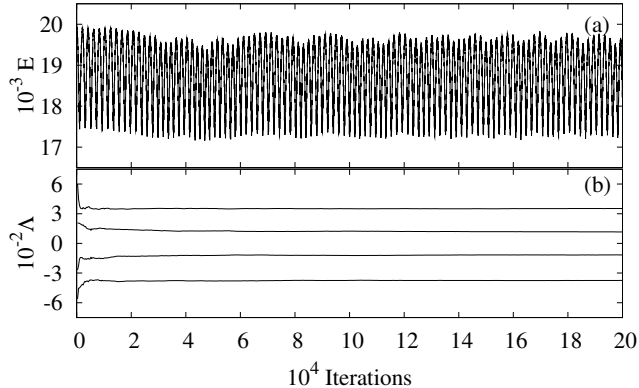}
 \end{center}
 \caption{(a) Energy from the system particle and (b) LE spectrum for $N=13$ 
    as a function of the iterations of the time series.}
  \label{LE-N10} 
  \end{figure}

Significant changes occur in this regime, where 
the time averaged energy starts to decay (see discussion in Section 
\ref{dissi}). We start by showing results for $N=13$. Figure 
\ref{LE-N10} shows the (a) energy from the system particle and (b) the four 
LEs as a function of time. Note that here we use just {\it one} trajectory 
and the time average of the particles energy is constant [See 
Fig.~\ref{LE-N10}(a)]). This is different from the result shown in
Fig.~\ref{En-N12-15} for $N=13$, where $800$ initial conditions were used. In 
Figure \ref{LE-N10}(b) we observe that two LEs are positive and two are 
negative. They are almost time independent. The sum of the LEs is zero 
meaning that the whole dynamics described by the TS of the system particle 
is conservative, as expected. The errors associated to the estimated LEs will 
be given later.
In Fig.~\ref{PS-N10}(b) the corresponding power spectrum for $N=13$ 
is shown. As in  Fig.~\ref{PS-N10}(a) we observe that the main peaks 
are well located at specific values of the fundamental frequency 
($\omega\sim0.56$Hz) and its high-harmonics. However, due to  
the $N=13$ oscillators, new smaller peaks appear 
close to the main frequencies. The power spectrum starts to show features
of broadness around the main frequencies. The power spectrum for $N=20$ is
shown in Fig.~\ref{PS-N10}(c). We see that almost all main frequencies
dissapear and a complicated spectrum is obtained. A chaotic dynamics is 
therefore expected for $N=20$. This result is confirmed with the LEs 
discussed next. 

\begin{figure}[htb]
 \begin{center}
\includegraphics*[width=8.5cm,angle=0]{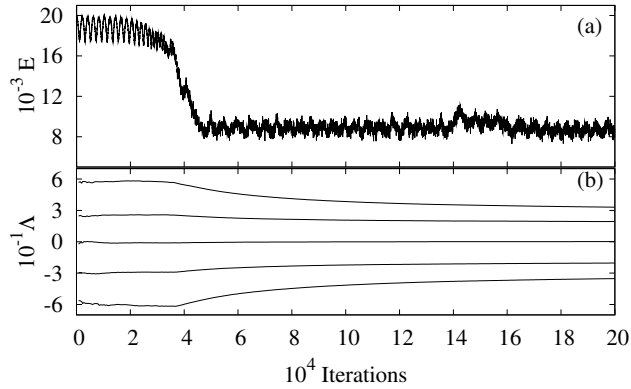}
 \end{center}
 \caption{(a) Energy for the system particle and (b) LE spectrum  
 as a function of the iterations of the time series and for $N=20$.}
  \label{LE-N15} 
  \end{figure}

Figure \ref{LE-N15} shows the (a) energy of the system particle and (b) 
the LEs as a function of time for $N=20$. As time increases, both positive 
LEs decrease when the oscillation {\it amplitude} of the energy 
decreases. In fact, each time the amplitude of the oscillation of energy 
decreases, energy starts to be transferred to the environment and the 
positive LEs decrease in time. Such local decreasing of the LEs in conservative
systems is due to ``sticky'' (trapped) trajectories \cite{zas02} which 
may also occur in higher-dimensional systems \cite{donnay,cesar1}. ``Sticky''
motion occurs when chaotic trajectories are trapped for a while in a
smaller portion of the phase space which is close to regular islands. 
When the trapped motion occurs in just one of the degrees of freedom, part 
of the total energy is transferred to the other degrees of freedom.
This is exactly what happens in Fig.~\ref{PS-N12-15} for
$N>13$. Part of the system energy is transferred to the environment and 
the particle begins to move in a smaller portion of the phase space (inner 
black curves). In Fig.~\ref{LE-N15} for example, at iterations 
$3\times 10^4\rightarrow 5\times 10^4$ part of the system particle energy 
is transferred to the environment and the local LEs start to decrease 
meaning that the chaotic trajectory of the system particles is approaching 
one (or more) regular islands. These regular islands must be related to the
motion of one (or more) of the $20$ harmonic oscillators. We note that 
in such a $2N+2$ high-dimensional system the trajectories may 
``penetrate''  the regular island due to the Arnold diffusion. For later 
iterations ($>5\times 10^4$)
the system particle continues to moves chaotically (trapped) around these 
regular islands. As $N$ increases, the regular islands are broken and the 
chaotic trajectory may find another regular island which will decrease its 
LEs.

\subsection{$N>20$}  
\label{N60-120}

\begin{figure}[htb]
 \begin{center}
\includegraphics*[width=9cm,angle=0]{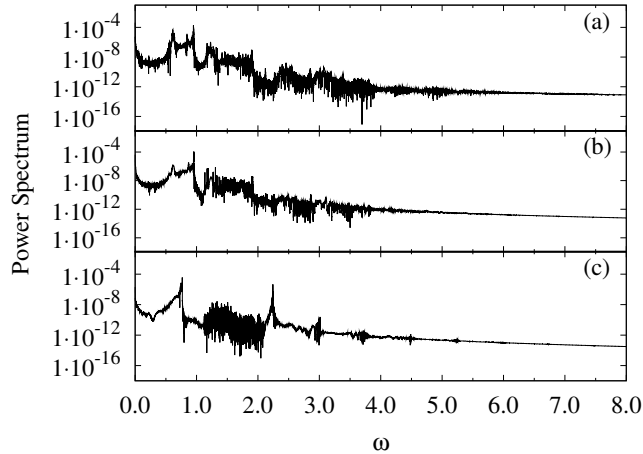}
 \end{center}
 \caption{Power spectra for (a) $N=60$, (b) $N=150$ and (c) $N=4000$.}
  \label{PS-N60} 
  \end{figure}

For higher values of $N$ the time dependence of the LEs
is not significantly different from the discussion above. Basically each 
time the the system particle energy starts to be transferred to the 
environment, the LEs decrease. As mentioned before, when $N$ increases
the system energy decay occurs faster and for shorter 
times. Figure \ref{PS-N60} shows the power spectra for $N=60,150,4000$.
Clearly it is possible to observe the complexity induced by the 
environment oscillators. The main frequencies observed at low 
values of $N$ [see Fig.~\ref{PS-N10}(a)] are now mixed to other (new) 
frequencies which arise due to the particle collisions with the 
oscillators. In particular, 
the main fundamental frequency at $w\sim0.56$ disappears. Higher
frequencies were excited and the power spectra indicate the chaotic 
dynamics for $N=60$ and $N=150$. For $N=4000$ the power spectrum  
shows  characteristics of a regular motion. In 
Figs.~\ref{PS-N60}(a)-(b) it is also possible to observe the new 
main frequency close to $w\sim0.98$ (For $N=4000$ this frequency is 
close to $\sim0.76$). The physical origin of these new frequencies 
will be discussed in the next paragraph. The case $N=4000$ 
[Fig.~\ref{PS-N60}(c)] shows very nicely what happens with the system 
particle. Besides the main fundamental frequency close to $w\sim0.76$, 
and its high-harmonics at $1.52,2.28,3.01\ldots$, all the oscillators 
frequencies in the interval $1.1\le w\le2.1$ can be clearly observed 
in the power spectrum. Note that the peak from the first high-harmonic 
at $w\sim1.52$ disappears inside the oscillators frequencies.

\begin{figure}[htb]
 \begin{center}
\includegraphics*[width=9cm,angle=0]{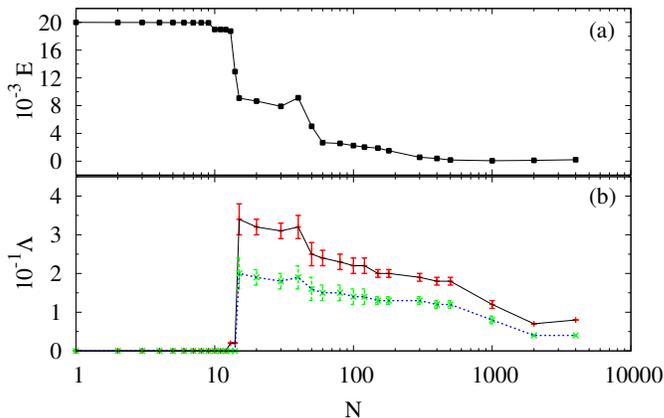}
 \end{center}
 \caption{(a) Final time average energy from the system particle and (b) 
      positive LEs as a function of $N$.}
  \label{LE-N} 
  \end{figure}

Next we give an overview of the behaviour of the system energy 
and the LEs as a function of $N$. Figure \ref{LE-N} shows the (a) final 
time average energy from the system particle and (b) positive estimated 
LEs as a function of $N$. To obtain the final time average energy, we 
took the time average of the energy over the last $2\times10^{3}$ 
integrated times. This is the final average
energy of the system particle before the simulation was stopped. It should 
not be confused with the initial conditions average energy from Section 
\ref{dissi}. We observe that for values of 
$N\lesssim12$, the initial energy ($\sim 0.02$) equals the final energy.
In other words, it is not transferred to the environment. For these values 
of $N$ the LEs are close to zero and the system dynamics is regular.
The errors for the estimated LEs are  shown as vertical bars. Since for 
$N\le12$ the embedding dimension is $m=2,3$, we have only one positive LE
($\sim10^{-3}>0$). Close to  $N\sim13,14$ the final mean energy starts to 
decrease, meaning that a portion of the system energy is transferred to 
the environment. Close to these values of $N$, where the embedding 
dimension is $m=4$,
both positive LEs increase and the system particle starts to behave 
chaotically. For higher values of $N$ the LEs decrease slowly 
[Fig.~\ref{LE-N}(b)] following the qualitative behavior of the mean 
final energy [Fig.~\ref{LE-N}(a)]. It is interesting to observe that very 
close to the interval of $N$ where the time 
averaged energy decay emerges ($N=13\rightarrow15$), both LEs increase 
and the system dynamics suffer a transition from a regular to a chaotic
motion. For higher values of $N$ the final mean energy of the system particle 
decreases, the LEs decrease, and the motion start to be regular again.
This can be seen by comparing Fig.~\ref{PS-N60}(c) with Fig.~\ref{LE-N} for
$N=4000$, for which the mean final energy is very small and the LEs are close 
to zero. Since the final energy of the system particle is very small, it will 
perform small oscillations around the minimum of the potential with a 
frequency $\omega_0$, which is the frequency of the linear motion around the 
minimum of the potential. Therefore, this harmonic motion induces the regular
dynamics. Since all frequencies in this paper are given in 
units of $\omega_0=1.0$, we conclude that for high values of $N$ the system 
particle oscillate very close to the minimum of the potential with the main 
frequency close to $w\sim\omega_0=1.0$. This is the physical origin of the
new frequencies which appear in Fig.~\ref{PS-N60}(c). 

From the LEs it is also possible to determine the Kaplan-Yorke dimension 
($D_{KY}$). Table \ref{TDK} shows $D_{KY}$ in the interval $1\le N\le 12$ with 
the corresponding values of $m,~\Lambda_1,~\Lambda_2$ and $\Lambda_3$ (for
$m=3$). In the interval $1\le N\le4$ we obtained $D_{KY}\sim 1.0$, 
for $5\le N\le12$ it is $D_{KY}\sim 2.2$, for $N=13$ it is  $D_{KY}\sim3.94$ 
and for $N=14$ it is $D_{KY}\sim3.88$, showing the fractal dimension of the 
reconstructed attractor and the dissipative behavior of the system particle. 
For higher values of $N$ we obtained approximately $D_{KY}\sim4.99$.

\begin{table}[!htb] 
  \centering
\caption{Kaplan-Yorke dimension ($D_{KY}$) in the interval $1\le N\le 12$ 
with the corresponding Lyapunov spectrum and values of $m$.}
\label{TDK}
    \large
    \begin{tabular}{|c|c|c|c|c|c|}
       \hline       
       $N$ & $m$ & $\Lambda_1$ & $\Lambda_2$ & $\Lambda_3$ & $D_{KY}$ \\
       \hline       
       1  & 2 & 0.0001 & -0.0023 & ***** & 1.00   \\
       \hline
       2  & 2 & 0.0001 & -0.0013 & ***** & 1.01   \\
       \hline
       3  & 2 & 0.0002 & -0.0035 & ***** & 1.01   \\
       \hline
       4  & 2 & 0.0012 & -0.0057 & ***** & 1.21   \\
       \hline       
       5  & 3 & 0.0078 & -0.0032 & -0.0190 & 2.21 \\
       \hline
       6  & 3 & 0.0100 & -0.0043 & -0.0350 & 2.16 \\
       \hline
       7  & 3 & 0.0097 & -0.0042 & -0.0215 & 2.26 \\
       \hline
       8  & 3 & 0.0103 & -0.0029 & -0.0261 & 2.28 \\
       \hline
       9  & 3 & 0.0095 & -0.0032 & -0.0271 & 2.23 \\
       \hline
       10 & 3 & 0.0096 & -0.0047 & -0.0255 & 2.19 \\
       \hline
       11 & 3 & 0.0098 & -0.0048 & -0.0265 & 2.19 \\
       \hline
       12 & 3 & 0.0115 & -0.0048 & -0.0273 & 2.25 \\
       \hline
    \end{tabular}
\end{table}

The main physical process responsible for the regular-chaotic transition 
occurs due to the overlap of resonances surfaces in the $2N+1$ (constant 
energy) dimensional space. As the consequence of the KAM 
(Kolmogorov-Arnold-Moser) theorem 
\cite{lichtenberg}, it is known that the breakup of tori occurs more 
likely at resonances between the tori of the unperturbed problem. 
For $N+1\ge3$  degrees of freedom, such resonance surfaces are not 
isolated from each other by the KAM surfaces and they may intersect in the 
constant energy surface, generating the Arnold web. As $N$ increases, the 
density of resonance surfaces increases and start to overlap so that 
a small coupling between the system and environment (the perturbation) is
strong enough to breakup the resonant tori, leading to the chaotic motion. 
This is the reason of the regular-chaotic transition as $N$ increases.
Consequently, the motion in Fig.~\ref{PS-N12-15}  (for $N\gtrsim13$)
comes from the combined effect of chaotic diffusion of trapped trajectories 
and Arnold diffusion. In Fig.~\ref{PS9} ($N=9$) and Fig.~\ref{PS15} 
($N=15$) we observe what happens in the phase-space of the environment 
oscillators at the regular-chaotic transition. In Fig.~\ref{PS9} all 
curves are closed giving the impression of a regular motion. For $N=15$ 
all curves start to display some irregularity. The most perturbed curves 
occur for $w_5$ and $w_8$, where the original dynamics in phase space 
(compare with Fig.~\ref{PS9}) is almost destroyed. These are the 
oscillators with lower frequencies but with larger amplitude of 
oscillations. Since the coupling between system and environment is 
bilinear (see Section \ref{model}), as the amplitude of the oscillators 
increases, the perturbation is larger and more likely to induce the 
chaotic motion (in addition to the overlap of resonances). 

Once the chaotic motion is reached, two distinct limit situations 
can be obtained as $N\rightarrow\infty$, total stochasticity or
regular motion \cite{lichtenberg}. In our case we clearly obtain 
the regular limit since the LEs decrease as $N\rightarrow 4000$
(confirmed by the power spectrum). As $N$ increases, due to the 
proximity of resonance surfaces,  more and more tori will be broken 
by the coupling. As a consequence, the trapped trajectory 
(see discussion related to Fig.~\ref{LE-N15}) will transfer energy
to the new degrees of freedom, the LEs will decrease more and more
so that for $N\rightarrow 4000$ they are almost zero. Besides that, 
in the Langevin description (for $N=\infty$), the system particle 
energy and the energy per oscillators approach zero, and a regular 
motion is expected. For a more detailed description of overlap of 
resonances in high-dimensional systems we refer the reader to 
\cite{lichtenberg}.

\section{Conclusions}
\label{conclusion}

In this paper we describe the nonlinear behavior of an open system 
(particle inside the anharmonic potential) interacting with an 
environment composed of a {\it finite} number $N$ of harmonic 
oscillators. The whole problem (System $+$ Environment 
$+$ Interaction) is conservative but, due to the energy exchange between 
system and environment, the {\it system only} is considered as an open 
system with dissipation. Since small dissipation is inevitable in real 
systems we analyse the nonlinear behavior when the time averaged energy
decay of the system particle starts to become relevant. The small dissipation
considered here is not modelled by a small damping constant, as usual, 
but by the finite number $N$ of oscillators. The microscopic dynamics as 
$N$ increases is not simple and its relation to the macroscopic 
statistical mechanics is even more complicated. 

To determine the values of $N$ where the time averaged energy decay 
emerges, we analyze the time evolution of the system particle mean energy, 
which is obtained over $800$ realizations of the environment variables. 
We show that the time averaged energy for the system particle starts to 
decay for $N$ in the interval ($10,20$). For lower values of $N$, system 
and environment just exchange energy and the time average of the mean 
particle energy remains constant (see Fig.~\ref{En-N12-15} for $N=12$). 
In the above mentioned interval, the system particle energy starts to be 
transferred to the environment (see Fig.~\ref{En-N12-15} for $N=13,14$). 
The transferred energy did not return back to the system for the whole 
range of integration time. As $N$ increases, the times 
for which the particles energy is transferred to the environment gets
smaller. From the nonlinear analysis we observed that the time series of $X(t)$ 
for a single trajectory of the system particle dynamics starts to be chaotic 
for $13\le N\le 15$. The power spectra confirm these results. This is 
exactly the interval of $N$ where the time averaged energy for the 
system particle starts to decay. 
For higher values of $N$ the final mean energy of the system particle 
decreases, the LEs decrease and the motion of the system particle starts to 
be regular again. This was nicely confirmed with the power spectrum for 
$N=4000$, shown in Fig.~\ref{PS-N60}(c). In this limit the LEs are close 
to zero and the system particle energy is very small. The particle  
oscillates close to the minimum of the anharmonic potential with a
frequency close to $w\sim\omega_0=1.0$, which is the frequency of the linear 
motion around the minimum of the potential. We also determined the 
Kaplan-Yorke dimension, which is  $\sim 1.0$ for $1\le N\le4$, 
 $\sim 2.2$ for $5\le N\le12$, $\sim 3.84$  for $13\le N\le 15$ and 
approaches $\sim 4.99$ for $N>15$. Interesting to mention here is that the 
Kaplan-Yorke dimension is fractal for $5\le N<15$, typical of dissipative
system. Therefore this dimension is not able to recognize that 
the whole dynamics is conservative. The origin of the chaotic motion is due 
to the overlap of resonances surfaces in the energy conserving surface as 
$N$ increases.

The above values of $N$ for which the time averaged energy decay and 
chaotic motion emerges, are valid for the specific oscillators frequencies 
generator used in the simulations.  These values may change a little if another 
generator is used \cite{Jane3}. However the qualitative behavior of all
analyzed quantities is the same when the time averaged energy starts to 
decay. 

Numerical evidences show a connection between the variation (in 
time) of the  {\it amplitude} of the particle energy with the energy decay 
and the decrease of the Lyapunov exponents. This is explained in terms
of chaotic trajectories from the system particles trapped close to regular
island. Since this trajectory is restricted to a smaller portion of the 
phase space, part of the total energy is transferred to other degrees of 
freedom thus explaining the origin of the time averaged energy decay.

\section*{Acknowledgments}
The authors thank CNPq and FINEP, under project CTINFRA-1, for financial 
support. C.~M.~thanks R.~L.~Viana and H.~A.~Albuquerque for helpful 
discussions.


\end{document}